\pdfoutput=1

\documentclass[twocolumn]{aastex62}
\usepackage{amsmath}
\usepackage{graphicx}
\usepackage{mathrsfs}
\usepackage{color}
\usepackage{url}
\usepackage{CJKutf8}
\usepackage{xcolor}

\received{September 21, 2019}
\revised{\today}
\accepted{2019}
\submitjournal{AJ}

\shorttitle{`Oumuamua}
\shortauthors{Hui \& Knight 2019}

\begin{document}

\title{
New Insights into Interstellar Object 1I/2017 U1 (`Oumuamua) from {\it SOHO}/{\it STEREO} Nondetections
}

\correspondingauthor{Man-To Hui}
\email{manto@ifa.hawaii.edu}

\author{
\begin{CJK}{UTF8}{bsmi}
Man-To Hui (許文韜)
\end{CJK}
}
\affiliation{Institute for Astronomy, University of Hawaii,
2680 Woodlawn Drive, Honolulu, HI 96822, USA}
\affiliation{Department of Earth, Planetary and Space Sciences, UCLA,
595 Charles Young Drive East,
Los Angeles, CA 90095-1567, USA}
\nocollaboration

\author{Matthew M. Knight}
\affiliation{Department of Astronomy, University of Maryland, 
College Park, MD 20742, USA}
\nocollaboration

\begin{abstract}

Object 1I/2017 U1 (`Oumuamua) is the first interstellar small body ever discovered in the solar system. By the time of discovery, it had already passed perihelion. To investigate the behavior of `Oumuamua around perihelion, we searched for it in {\it Solar and Heliospheric Observatory} ({\it SOHO}) and {\it Solar TErrestrial RElations Observatory} ({\it STEREO}) images from early 2017 September (preperihelion), but did not detect it. The nondetection of `Oumuamua by {\it STEREO} renders more stringent constraints on its physical properties thanks to the extreme forward-scattering observing geometry. Assuming geometric albedo $p_V = 0.1$, the effective scattering cross-section of any dust coma was $\la \left(2.1 \pm 0.2 \right) \times 10^{4}$ m$^{2}$. Assuming it behaved like a typical solar-system comet this would correspond to a total mass of $\la 20 \pm 2$ kg, and a water production rate of $\la \left(6.1 \pm 0.5 \right) \times 10^{25}$ s$^{-1}$ at heliocentric distance $r_{\rm H} = 0.375$ au. If scaled to post-discovery $r_{\rm H}$, the water production rate would be smaller than any of the previously reported upper limits by at least an order of magnitude. To exhibit the reported nongravitational motion with our default assumptions requires a nucleus bulk density $\la$40 kg m$^{-3}$; higher bulk densities are possible for other assumptions. Alternatively, we show that thermal fracturing could have plausibly removed an inert surface layer between these observations and discovery, thus initiating activity after `Oumuamua left the field of view of {\it STEREO}.

\end{abstract}


\section{Introduction}

Object 1I/2017 U1 (`Oumuamua) (formerly designated as C/2017 U1 then A/2017 U1 by the Minor Planet Center, hereafter `Oumuamua) was first discovered by the Panoramic Survey Telescope and Rapid Response System located at Haleakala, Hawaii, on 2017 October 18, and then was soon recognised as a small body from beyond the solar system because of the significantly hyperbolic eccentricity ($e = 1.201$) of the heliocentric orbit \citep{2017MPEC....U..181B,2017MPEC....U..183M}. It had passed perihelion by the time of discovery (perihelion epoch $t_{\rm p} = 2017$ September 9.5), and is the first interstellar interloper ever observed in the solar system \citep{2018A&A...610L..11D}.

An intense number of observations of `Oumuamua by various means were conducted immediately after the announcement of the discovery \citep{2017ApJ...851L..38B,2018ApJ...856L..21B,2018ApJ...852L...2B,2018NatAs...2..407D,2018NatAs...2..133F,2017ApJ...850L..36J,2017ApJ...851L..31K,2017arXiv171009977M,2017Natur.552..378M,2018Natur.559..223M,2018AJ....155..185P,2018AJ....156..261T,2017ApJ...851L...5Y}, based upon which we learn that `Oumuamua has an unremarkable colour compared to small bodies from the solar system, featureless visible and near-infrared spectra, and clear brightness variations indicative of an elongated shape and complex non-principal axis rotation \citep{2018NatAs...2..407D, 2018NatAs...2..383F, 2018ApJ...856L..21B,2019MNRAS.489.3003M}. An unambiguous detection of a nongravitational acceleration of `Oumuamua using postperihelion astrometric observations was reported by \citet{2018Natur.559..223M}. Despite the clear detection of nongravitational acceleration, morphology consistent with cometary outgassing was not observed in deep searches by many authors. Constraints on the mass-loss activity were derived from various observations of `Oumuamua \citep{2017ApJ...850L..36J,2017Natur.552..378M,2017ApJ...851L...5Y,2018AJ....155..185P,2018AJ....156..261T}. For instance, \citet{2017Natur.552..378M} estimated that the dust production rate was $\la2 \times 10^{-3}$ kg s$^{-1}$ at heliocentric distance $r_{\rm H} \approx 1.4$ au, while the production rate of OH at $r_{\rm H} = 1.8$ au was $\la 2 \times 10^{27}$ s$^{-1}$ \citep{2018AJ....155..185P}. Possible explanations include subsolar venting activity of volatiles \citep{2019ApJ...876L..26S}, that the object is an icy fractal aggregate \citep{2019ApJ...872L..32M},  preperihelion disintegration which turned `Oumuamua into a massive cloud of dust debris \citep{2019arXiv190108704S}, and even the solar sail hypothesis \citep{2018ApJ...868L...1B}. While all of these hypotheses were designed to produce large nongravitational accelerations, none actually directly fit their models to the astrometry and the source of the nongravitational acceleration remains uncertain.

Formation models generally prefer that `Oumuamua should have been more like a comet rather than an asteroid \citep[e.g.,][]{1989ApJ...346L.105M,2018ApJ...852L..15C,2018MNRAS.476.3031R,2018ApJ...856L...7R}. Based upon the observations of comets in our solar system, one would expect that the activity of `Oumuamua, if any, should be more pronounced around perihelion, and therefore easier to be observed. `Oumuamua was at solar elongations $\la$45\degr~-- the approximate limit of professional sky surveys -- from 2017 August 11 through September 30. As a result, the earliest precovery observations were on 2017 October 15, only five days earlier than the discovery images. Although the small solar elongations prevented serendipitous observations from ground-based telescopes, they were well suited for {\it Solar and Heliospheric Observatory} ({\it SOHO}) and {\it Solar TErrestrial RElations Observatory} ({\it STEREO}). In particular, `Oumuamua was in an extremely forward-scattering regime (large phase angles) from the perspective of {\it STEREO}, whereby dust around the nucleus would be enhanced in terms of apparent brightness by orders of magnitude. This mechanism has enabled a number of intrinsically mediocre comets to be visible even in broad daylight \citep[see][and citations therein]{2007ICQ....29...39M}, and has resulted in significant brightening of many comets observed by {\it SOHO} and {\it STEREO} \citep[e.g., ][]{2004A&A...427..755G,2010AJ....139..926K,2013MNRAS.436.1564H}. Furthermore, physical properties of cometary dust can be probed from the scattering behaviour at large phase angles \citep[e.g., ][]{2004come.book..577K,2013MNRAS.436.1564H}. 

We thus searched for potential {\it SOHO}/{\it STEREO} observations of `Oumuamua around perihelion, in a hope of extending the observed arc or using a nondetection to better constrain its physical properties, although we are well aware that the {\it SOHO}/{\it STEREO} images are generally not as deep or good-quality as the ground-based data of `Oumuamua.

The paper is structured in the following manner. We detail the {\it SOHO}/{\it STEREO} observations in Section \ref{sec_obs}, and present the results in Section \ref{sec_anls}. Discussion is held in Section \ref{sec_disc}, while the summaries of the analysis are in Section \ref{sec_sum}.

\begin{figure}
\gridline{\fig{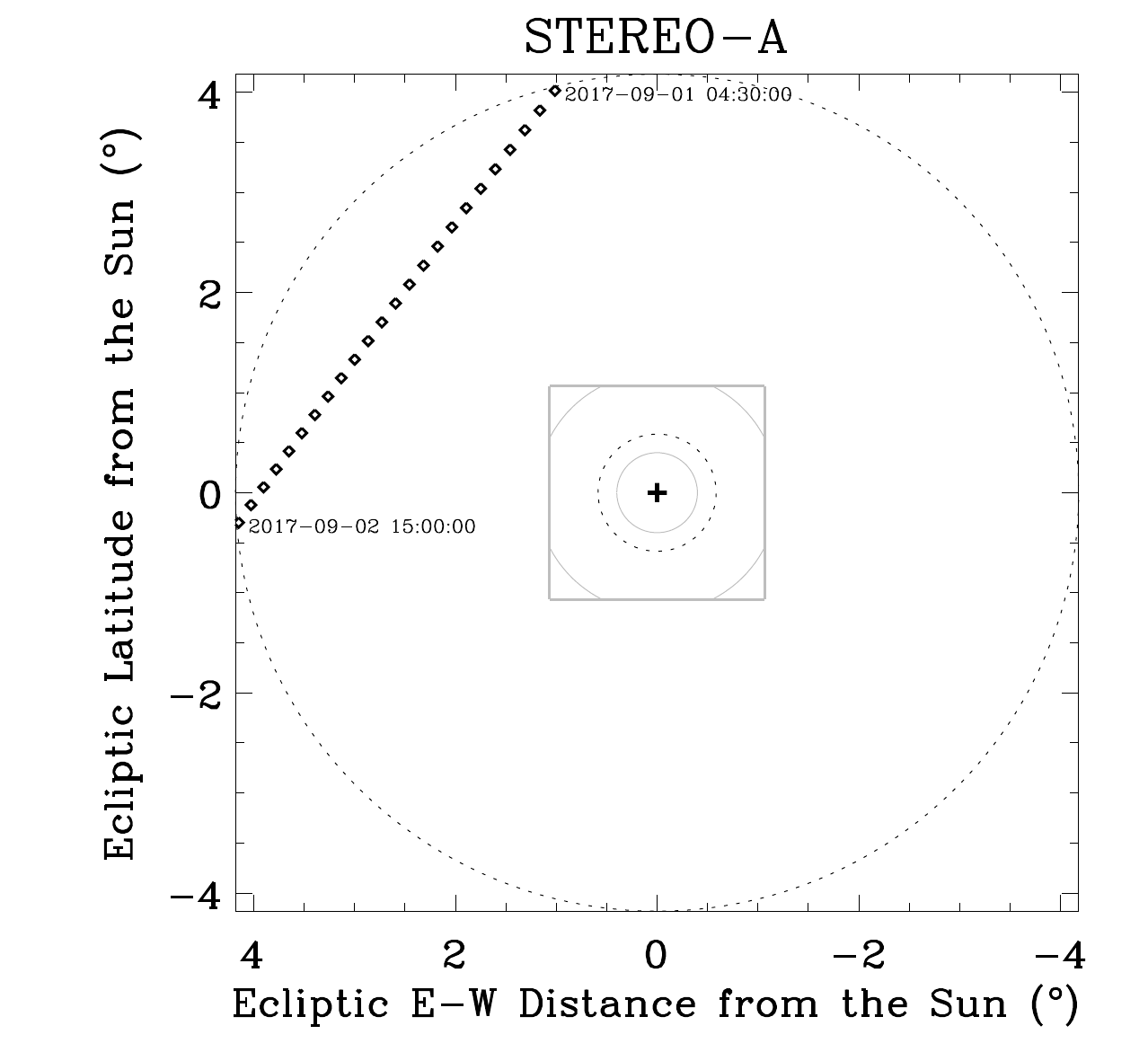}{0.5\textwidth}{(a)}
          }
\gridline{\fig{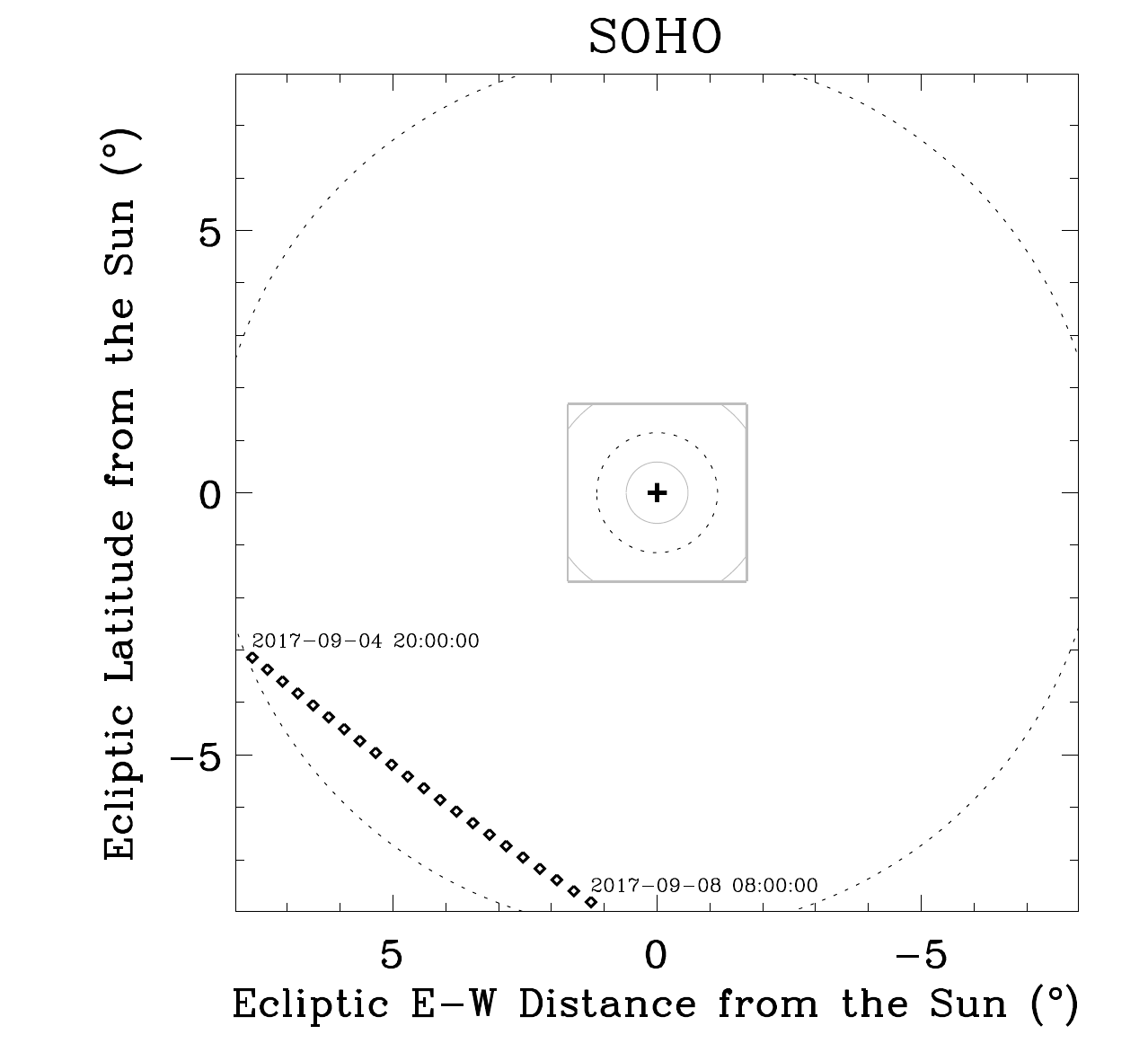}{0.5\textwidth}{(b)}
          }
\caption{
Apparent paths of 1I/2017 U1 (`Oumuamua) observed from (a) {\it STEREO-A} and (b) {\it SOHO}. The symbol ``+'' marks the centre of the Sun. In panel (a), the annulus between the dotted circles is the observable region of COR2, whereas the FOV of COR1, the other coronagraph onboard {\it STEREO}, is marked by the grey box around the Sun, with its observable region confined by two grey thin circles. In panel (b), the plot scheme is the same, but corresponds to coronagraphs C3 and C2, respectively. The positions are plotted every 90 min and 4 hr in panels (a) and (b), respectively. Times are in UTC.
\label{fig_path}
} 
\end{figure}

\begin{figure*}
\gridline{\fig{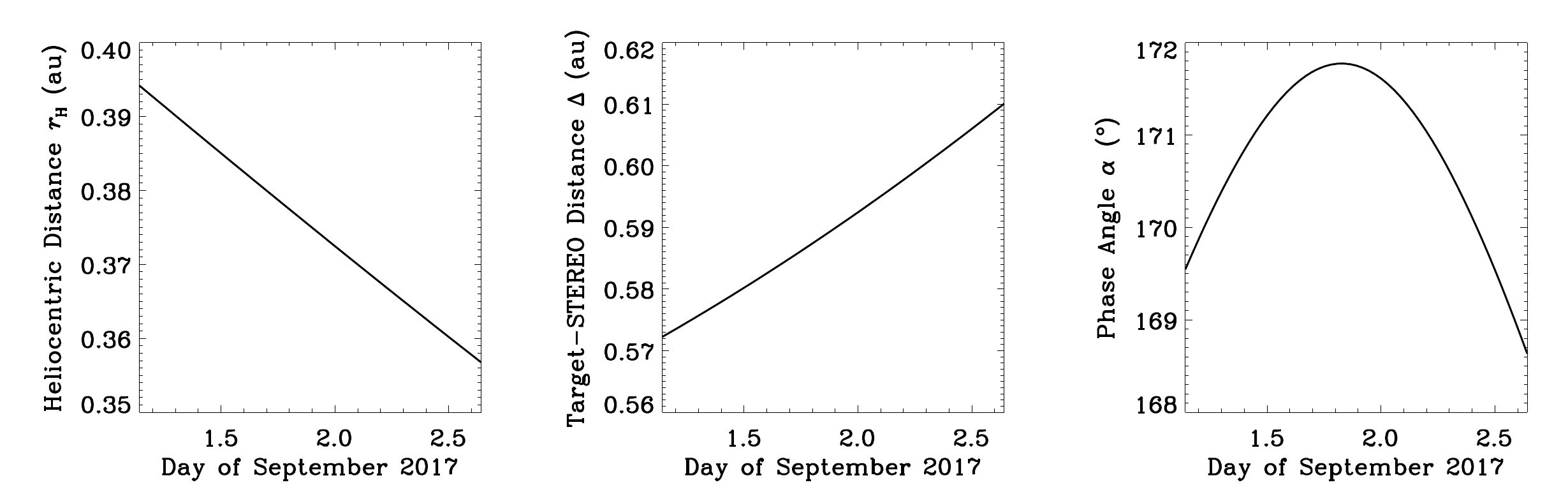}{1.0\textwidth}{(a)}
          }
\gridline{\fig{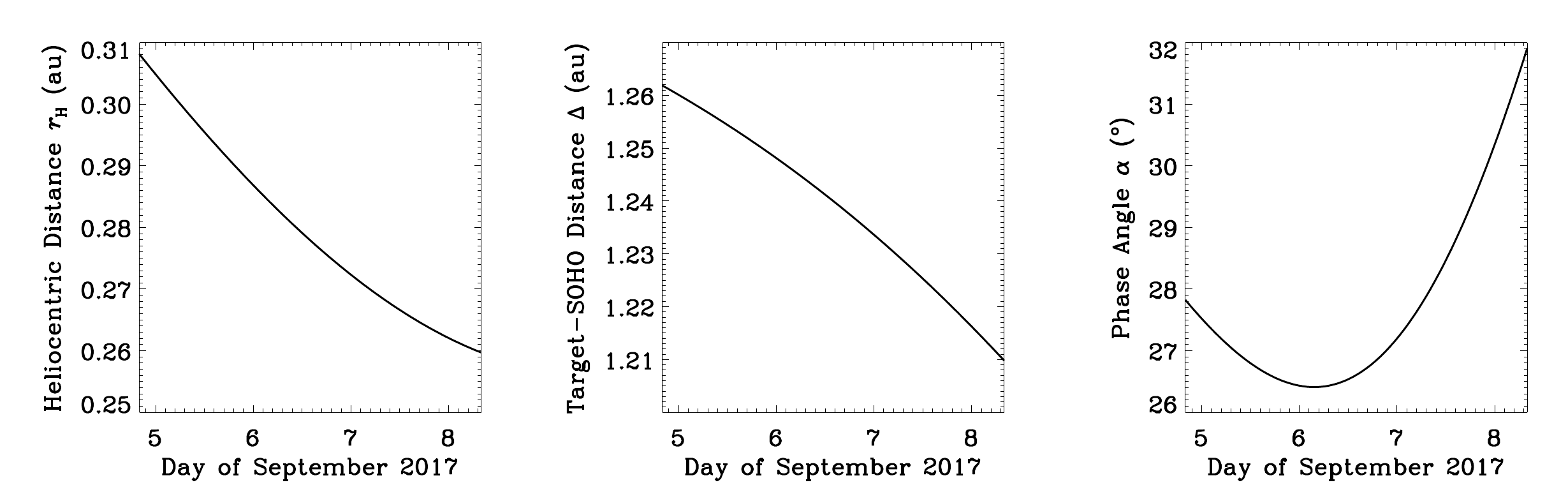}{1.0\textwidth}{(b)}
          }
\caption{
Observing geometry of 1I/2017 U1 (`Oumuamua) as functions of time from the perspectives of {\it STEREO-A} (a) and {\it SOHO} (b). Noteworthily, during the transit in {\it STEREO-A}, the interstellar interloper was in an extremely forward-scattering regime at phase angle $\alpha \gtrsim 169\degr$. Although `Oumuamua was closer to the Sun during its transit of {\it SOHO}, the distance from {\it SOHO} was much further than that from {\it STEREO-A}, and additionally, the phase angle was unimpressively ordinary, making the overall observing geometry from {\it SOHO} much inferior to that from {\it STEREO-A}. The perihelion epoch of `Oumuamua is $t_{\rm p} = 2017$ September 09.5.
\label{fig_vgeo}
} 
\end{figure*}

\section{Observations}
\label{sec_obs}

{\it SOHO} is a solar satellite that continuously monitors the Sun and its neighbourhood at the L1 Lagrangian point of the Sun and Earth system \citep{1995SoPh..162....1D}. In this study, we only focus on one of its onboard Large Angle and Spectrometric Coronagraph \citep{1995SoPh..162..357B} -- C3. Full-resolution images from the camera have a dimension of 1024 $\times$ 1024 pixel, a pixel scale of 56\farcs1 pixel$^{-1}$, and the fields-of-view (FOVs) are annular (3.7-30 $R_\odot$, where $R_\odot$ is the apparent solar radius, or 1\fdg0-8\fdg0). The camera uses four different filters (blue: 4200-5200 \AA, clear: 4000-8500 \AA, orange: 5400-6400 \AA, and red: 7300-8350 \AA), though we only consider the clear filter images because virtually all images were acquired in that bandpass and the other filters were all acquired at half-resolution.

{\it STEREO} consists of a pair of identical spacecraft orbiting around the Sun in Earth-like orbits, with one leading Earth ({\it STEREO-A}) and the other trailing Earth ({\it STEREO-B}) \citep{2005AdSpR..36.1483K}. Unfortunately, communication with {\it STEREO-B} were lost in 2014.\footnote{\url{https://stereo-ssc.nascom.nasa.gov/behind_status.shtml}} We only focus on total brightness images taken by one of the coronagraphs, COR2, onboard {\it STEREO-A} \citep[denoted as COR2-A;][]{2008SSRv..136...67H}, since it was the only camera which would have recorded `Oumuamua around perihelion. The camera uses 2048 $\times$ 2048 CCD arrays and continuously observes regions 2.5-15 $R_\odot$ (0\fdg7-4\fdg0) from the Sun, with a full-resolution image scale of 14\farcs7 pixel$^{-1}$ and a bandpass of 6500-7500 \AA.

We show the apparent trajectories of `Oumuamua in Figure \ref{fig_path} and the observing geometry in Figure \ref{fig_vgeo}, from {\it STEREO} (top) and {\it SOHO} (bottom). `Oumuamua entered the FOV of COR2-A on 2017 September 01 and stayed inside the FOV for nearly 35 hr before egress, during which a total of 107 images (each having equivalent exposure time of 6 s) were taken. It subsequently went into the FOV of C3 on 2017 September 04 and exited on September 08, during which a total of 360 images (each having equivalent exposure time of 20 s) were taken.


We processed the publicly available level-0.5 {\it SOHO} and {\it STEREO} FITS images by following similar procedures as \citet{2010AJ....139..926K} using the IDL SolarSoftWare library \citep{1998SoPh..182..497F}. For each bias subtracted and flat-fielded image, a median background was computed from neighbouring images with exactly the same configurations (filter, polariser, binning) and was then subtracted. In this way, the majority of the corona, which would be otherwise dominant, was removed. By visual inspection, we found no noticeable artefacts that were introduced in the aforementioned step or attributable to variations in solar activity in the final images. 

\begin{figure*}
\epsscale{1.}
\begin{center}
\plottwo{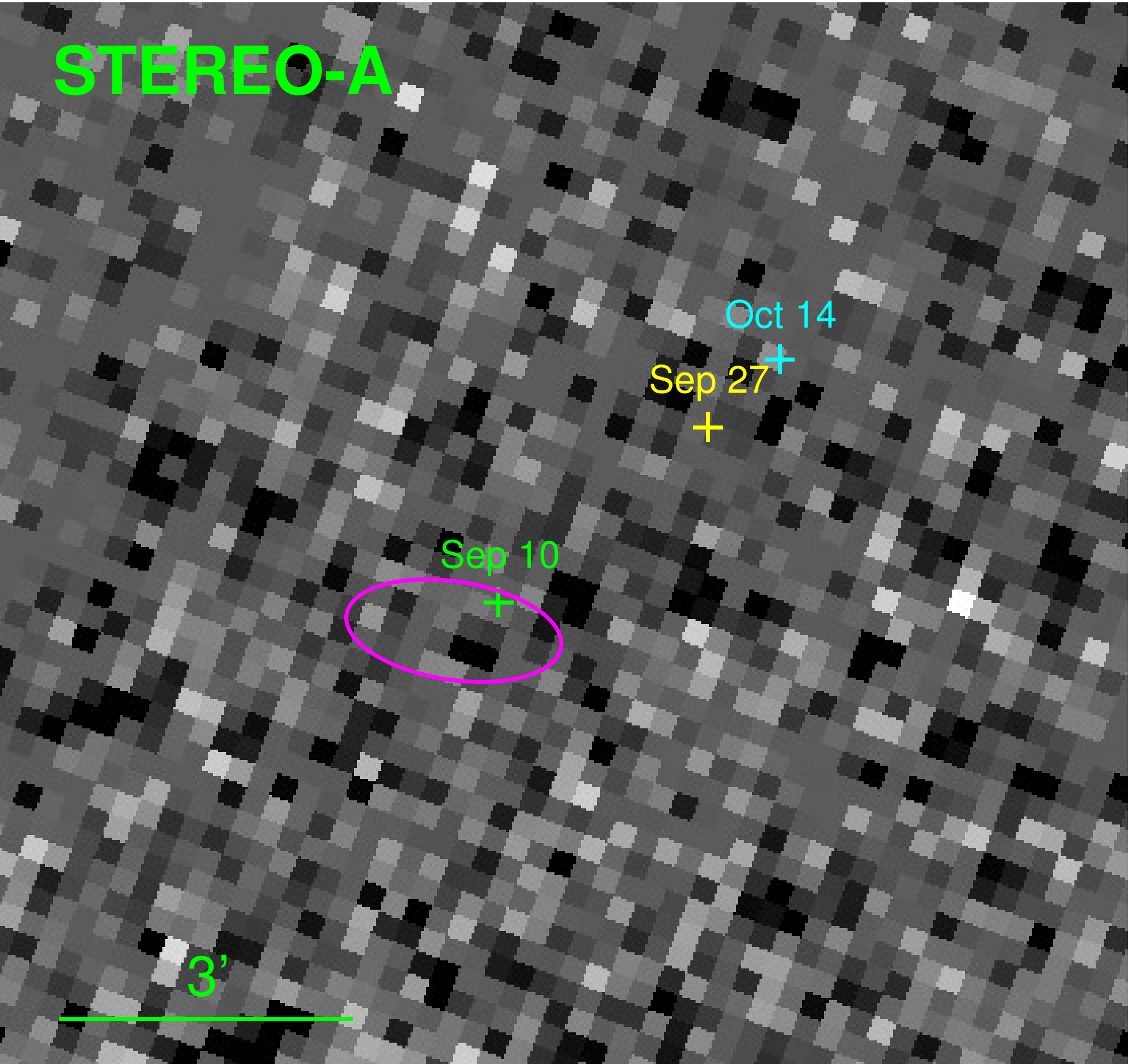}{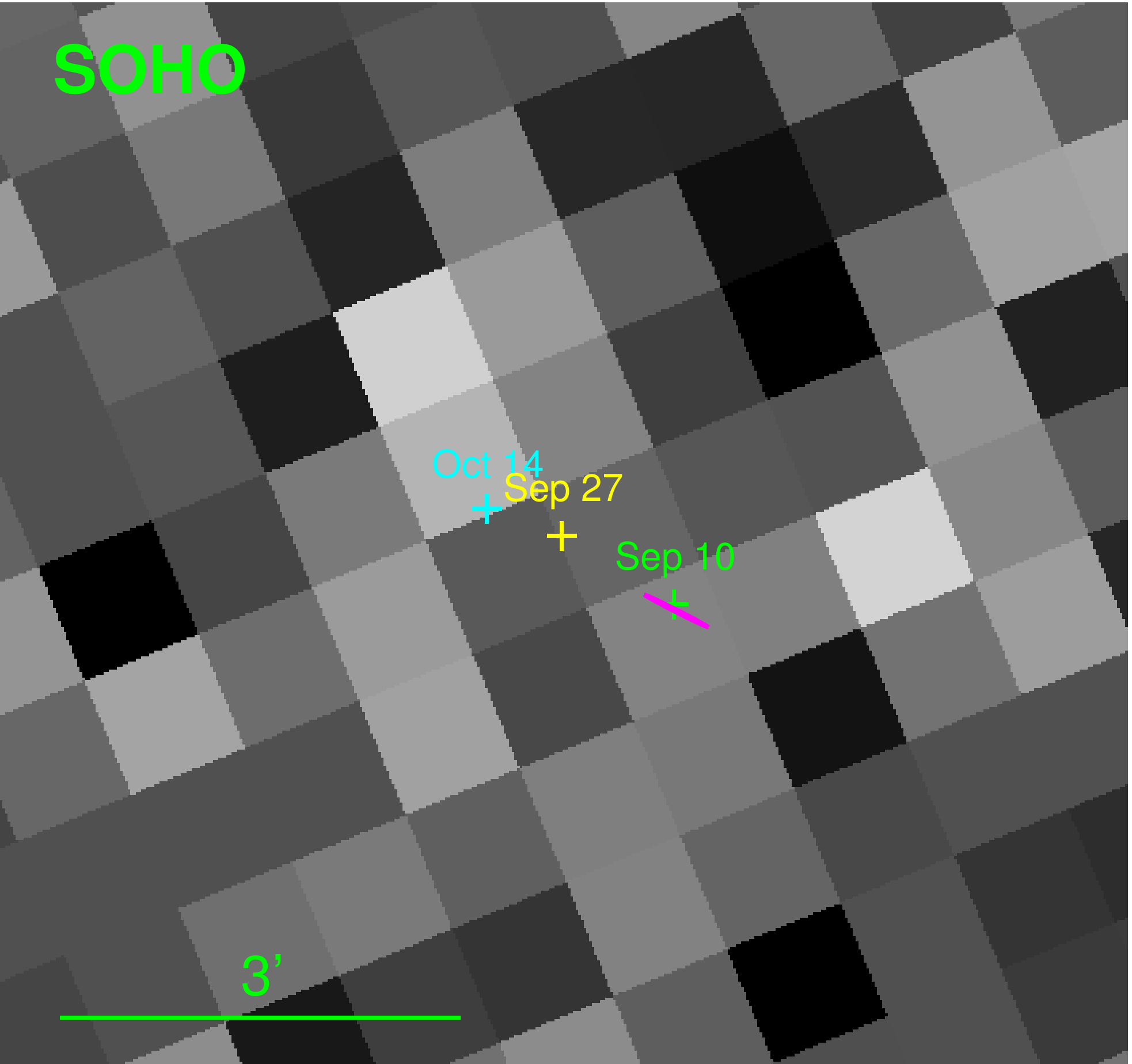}
\caption{
Coadded {\it STEREO-A} (left) and {\it SOHO} (right) images around the predicted positions of `Oumuamua. The magenta ellipse in the panels is the $3\sigma$ positional uncertainty ellipse on the sky plane based on the nongravitational solution by JPL Horizons at referenced epochs 2017 September 01 20:24 UT for {\it STEREO} and September 06 10:30 UT for {\it SOHO}. The green, yellow and cyan crosses correspond to the positions of `Oumuamua in the three different nongravitation models, with $t_{\rm NG}$ labelled (all in year 2017). Equatorial north is up and east is left.
\label{fig_img}
}
\end{center} 
\end{figure*}

\section{Analysis}
\label{sec_anls}

With the ephemeris calculated by JPL Horizons based on the nongravitational solution by \citet{2018Natur.559..223M}, we did not detect `Oumuamua in any of the individual {\it SOHO} or {\it STEREO} images. The $3\sigma$ positional uncertainties are small, corresponding to $\lesssim$5 pixels ($\sim$1\farcm1) for COR2-A images, and less than one pixel ($\sim$0\farcm3) for C3 images. This is due to the low resolution of the coronagraph images.

We were aware of the possibility that the nongravitational force of `Oumuamua may have changed during the period between its transit in the FOVs of {\it SOHO}/{\it STEREO} cameras and the discovery, which is unaccounted in the orbital solution. To see how this would affect the uncertainty region, we assumed the nongravitational force of `Oumuamua to be a step function of time: `Oumuamua did not exhibit the nongravitational effect until some specific epoch $t_{\rm NG}$. We tried three different epochs: 2017 September 10.0, 27.0 and October 14.0 (in units of the internal Barycentric Dynamical Time, i.e., TDB), all of which are earlier than the earliest astrometry fit by \citet{2018Natur.559..223M} and could therefore fit the available astrometric observations. The largest positional deviation from the nominal orbit is from the orbit with $t_{\rm NG} = $~TDB 2017 October 14.0 ($\sim-4\arcmin$ in RA, $\sim+3\arcmin$ in Decl. in {\it STEREO-A}; $\sim+2\arcmin$ in RA, $\sim+1\arcmin$ in Decl. in {\it SOHO}). However, because of the low resolution of the spacecraft, this would only correspond to $\la$20 pixels in COR2-A images, and $\la$2 pixels in C3 images at worst; the different nongravitation models give similar apparent motion rates during the {\it SOHO}/{\it STEREO} observations and do not introduce noticeable differences in the coadded images (discussed in the next paragraph). We thereby conclude that the orbital uncertainty of `Oumuamua should have a negligible effect on our detection attempt.

In order to boost the signal-to-noise ratio (SNR) of `Oumuamua, we also coadded the images with registration on the expected apparent motion rate of the nominal orbit. For COR2-A images, we grouped the image sequence into several groups. Within each group, a median-combined image was computed. However, we found no hint of `Oumuamua around the expected positions whatsoever. Finally we even coadded the whole image sequence into a single average/median image using the ephemeris of `Oumuamua, but the result remains the same that nothing shows up above the background noise at the expected position (Figure \ref{fig_img}). We thus conclude that it is most likely that `Oumuamua was too faint to be visible in the {\it STEREO} images near perihelion.

In order to quantify the apparent faintness of `Oumuamua during the {\it STEREO} observations, we coadded the images in the exact same way but with registration on the field stars in AstroImageJ \citep{2013ascl.soft09001C,2017AJ....153...77C}. In order to determine the zero-point of the coadded image, we first employed Sextractor \citep{1996A&AS..117..393B} to pick up field stars with SNR $\ge 3$. Candidates within 500 pixels of the image center were discarded because the noise due to the uncleaned corona near the occulter was too high. We were left with $\sim$850 field stars. Then we queried the AAVSO Photometric All-Sky Survey Data Release 9 \citep[APASS-DR9;][]{2016yCat.2336....0H} catalog for all stars within 6\degr~of the center of the FOV with apparent {\it V}-band magnitudes $m_{\ast,V} < 12$ and having magnitudes available in all five filters: {\it B}, {\it V}, {\it g}', {\it r}', and {\it i}'. The candidate sources were matched with the catalog stars. We then performed a linear fit to the zero-point and some color index $C$ (including {\it B} $-$ {\it V}, {\it g}' $-$ {\it r}', {\it r}' $-$ {\it i}', and {\it V} $-$ {\it r}'), and thereby obtained a best-fit function in the following form:
\begin{equation}
m_{V} = ZP_{0} - 2.5 \log \mathscr{F} + k_{\rm C} C
\label{eq_ZP},
\end{equation}
\noindent where $ZP_{0}$ is the zero-point at color index $C = 0$, $\mathscr{F}$ is the integrated flux, and $k_{C}$ is the corresponding slope. We measured the sky background uncertainty around `Oumuamua. At a $3\sigma$ level, given the color of `Oumuamua \citep[e.g.,][]{2017ApJ...850L..36J}, we obtained the apparent magnitude of `Oumuamua to be $m_V \ga 13.0 \pm 0.1$, where the uncertainty is the standard deviation of $m_V$ from different color-index tests.

For the {\it SOHO} data, we adopted a different way, as the field distortion of the C3 images are so strong that AstroImageJ could not fully cope: the stars in the resulting coadded image with registration on the field stars were not aligned properly and appear conspicuously trailed except near the image center. Therefore, we decided to examine the C3 images when asteroids (2) Pallas, (4) Vesta, and (10) Hygiea were in the FOV. Without any difficulty, we spotted Vesta at $m_V = 7.8$ (JPL Horizons) in the image coadded from a daily sequence from 2017 September 27 (122 images). Pallas at $m_V = 9.9$ (JPL Horizons) was barely visible (SNR $= 1.6$) in a stack of all the 100 image taken from 2017 March 06. We could not recover Hygiea at $m_V = 11.1$ (JPL Horizons) in a full day stack from 2018 February 03 (108 images). Thus, we estimate that the limiting magnitude for a full day stack is likely $m_V \approx 9.9$. Given that we coadded a total number of 358 {\it SOHO} images on `Oumuamua (two images in which `Oumuamua was too close to the edge were discarded), we expect the corresponding limiting magnitude of the stack to be $m_{V} \approx 10.6$.

\section{Discussion}
\label{sec_disc}

We now attempt to constrain the activity of `Oumuamua using our {\it SOHO}/{\it STEREO} nondetections of the object. For the sake of calculations, we assume that `Oumuamua behaves like a typical solar-system comet. If `Oumuamua was completely inactive during its transits in {\it SOHO} and {\it STEREO}, its apparent magnitude would be $m_{V} \ga 21$ and $\ga 35$, respectively (JPL Horizons), which are well below the detection limits of the two spacecraft. In fact, the only inactive objects observed by {\it SOHO}/{\it STEREO} are the largest main-belt asteroids; all other objects are observed due to the large scattering cross-sections of their dust comae \citep{2018SSRv..214...20J}. Although `Oumuamua was slightly closer to the Sun when viewed by {\it SOHO} (left panels in Figure \ref{fig_vgeo}), it was closer to {\it STEREO} (middle panels in Figure \ref{fig_vgeo}) and, critically, was at an extremely large phase angle ($\alpha \ga 169\degr$) when viewed by {\it STEREO} (right panels in Figure \ref{fig_vgeo}). If there was a dust coma around the nucleus with physical properties similar to those of solar-system comets, we would expect to witness a strong forward-scattering effect whereby the apparent brightness of the target would be intensified by two orders of magnitude when observed by {\it STEREO} \citep{2007ICQ....29...39M,2013MNRAS.436.1564H}, with minimal brightness increase expected for the {\it SOHO} phase angle. Thus, the {\it STEREO} nondetection of `Oumuamua should render us a much more stringent physical constraint of the target, and so in the following paragraphs we focus on the {\it STEREO} result only.

Assuming that `Oumuamua had prominent mass-loss activity around perihelion such that the ejected dust would dominate the scattering, the effective scattering geometric cross-section, denoted as $C_{\rm e}$, is related to the apparent magnitude $m_V$ by the following equation:
\begin{equation}
C_{\rm e} = \frac{\pi r_{\rm H}^2 {\it\Delta}^2}{p_{V} \phi\left(\alpha\right)r_{\oplus}^{2}} 10^{0.4 \left(m_{\odot, V} - m_{V} \right)}
\label{eq_xs},
\end{equation}
\noindent where $r_{\rm H}$ and ${\it \Delta}$ are the heliocentric and target-observer distances, respectively, $r_{\oplus} = 1~{\rm au} \approx 1.5 \times 10^{8}$ km is the mean Sun-Earth distance, $p_V$ is the {\it V}-band geometric albedo, $\phi \left(\alpha\right)$ is the phase function of scattering dust \citep[e.g.,][]{2011AJ....141..177S}, and $m_{\odot,V} = -26.74$ is the apparent {\it V}-band magnitude of the Sun observed at $r_{\rm H} = r_{\oplus}$. Plugging time-average $r_{\rm H} = 0.375$ au, ${\it \Delta} = 0.590$ au, and $\alpha = 170\fdg8$ along with a nominal albedo of $p_V = 0.1$ into Equation (\ref{eq_xs}), we obtain $C_{\rm e} \la \left(2.1 \pm 0.2\right) \times 10^4$ m$^2$ for `Oumuamua.

Now, let us further assume that the dust coma would consist of dust grains with mean radius $\bar{\mathfrak{a}} \approx 0.7$ \micron~(corresponding to the peak transmissivity wavelength of the COR2 camera) and a bulk density of $\rho_{\rm d} = 1$ g cm$^{-3}$. Then the total mass of the dust coma of `Oumuamua would be $\mathcal{M}_{\rm d} = 4\rho_{\rm d} C_{\rm e} \bar{\mathfrak{a}}/3 \la 20 \pm 2$ kg, which is many orders of magnitude smaller than the estimated masses of cometary debris clouds after known disintegration events of sub-kilometer or smaller comets \citep[e.g.,][]{2015ApJ...813...73H,2019arXiv190901964J}. The only way to boost $\mathcal{M}_{\rm d}$ by a few orders of magnitude is that the dominant dust of the coma of `Oumuamua is much larger than our assumed value by a factor of $\ga$$10^{2}$ (i.e., dominant grain radius $\bar{\mathfrak{a}} \ga 0.1$ mm). Such larger dust grains are not constrained by other investigations of `Oumuamua \citep[e.g., the meteor search by ][]{2017ApJ...851L...5Y}, but if they existed, the forward-scattering effect would be much more concentrated at phase angle $\alpha = 180\degr$ and the brightness surge would be significantly steeper \citep{1998asls.book.....B}. However, never have we observed any solar-system comets exhibiting such a steep forward-scattering enhancement \citep[e.g., ][]{2004come.book..577K,2007ICQ....29...39M}. Nor have we observed any comets with dominant grains of $\bar{\mathfrak{a}} \ga 0.1$ mm in radius in the inner solar system \citep[see][]{2004come.book..565F}. Therefore, we think that the {\it STEREO} nondetection likely rules out the hypothesis by \citet{2019arXiv190108704S} that `Oumuamua had been a debris cloud of $\sim$$10^{4}$ kg in mass by the time of these preperihelion observations since such a cloud should have been easily visible in {\it STEREO} images.

Unfortunately, the low resolution of the {\it STEREO} images inhibits us from using a method in which we assume that the dust coma is distributed fully in the photometric aperture to estimate the corresponding mass-loss rate of `Oumuamua. Instead, we exploit the empirical correlation by \citet{2008LPICo1405.8046J} to estimate the water production rate of `Oumuamua to be $Q \la \left(6.1 \pm 0.5\right) \times 10^{25}$ s$^{-1}$ from the minimum apparent magnitude reduced to ${\it \Delta} =1$ au by the inverse square law. We understand that none of the comets sampled by \citet{2008LPICo1405.8046J} are at the same reduced brightness level of `Oumuamua; however, this correlation appears to remain valid for objects with very low production rates such as active asteroid P/2016 J1 (PANSTARRS) \citep{2017AJ....153..141H}. Assuming the dust-to-gas mass ratio $\mathcal{X} = 1$, we obtain the dust mass-loss rate of `Oumuamua as $\left \langle \dot{\mathcal{M}}_{\rm d} \right \rangle \la 1.8 \pm 0.2$ kg s$^{-1}$.

We then proceed to numerically solve the energy conservation equation
\begin{equation}
\left(1-A_{\rm B}\right) S_\odot \left(\frac{r_{\oplus}}{r_{\rm H}} \right)^2 \cos \zeta = \epsilon \sigma T^4 + L\left(T\right) f_{\rm s}
\label{eq_E}
\end{equation}
\noindent for the mass flux of water $f_{\rm s}$ during the {\it STEREO} observations. Here, $A_{\rm B}$ is the Bond albedo, $S_\odot = 1361$ W m$^{-2}$ is the solar constant, $1/4 \le \cos \zeta \le 1$ is the illumination efficiency (the lower boundary corresponds to an isothermal nucleus, and the upper one corresponds to the subsolar scenario), $\epsilon$ is the emissivity, $\sigma = 5.67 \times 10^{-8}$ W m$^{-2}$ K$^{-4}$ is the Stefan-Boltzmann constant, and $L$ is the latent heat of water ice as a function of the surface temperature $T$. We assign $A_{\rm B} = 0.04$ and $\epsilon = 0.9$, which are both typical values for comets and asteroids in the solar system \citep[e.g., ][]{2004come.book..223L,2005Icar..173..153L,2016ApJ...817L..22L}. Substituting into Equation (\ref{eq_E}) yields $8 \times 10^{-4} \la f_{\rm s} \la 3 \times 10^{-3}$ kg m$^{-2}$ s$^{-1}$. Therefore, the active surface area of `Oumuamua likely did not exceed $\sim$$2 \times 10^3$ m$^{2}$. Compared to the total surface area of `Oumuamua \citep[$\sim$$8 \times 10^4$ m$^2$, assuming a prolate spheroid with semimajor axes $\mathcal{R}_1 \approx 230$ m, $\mathcal{R}_2 = \mathcal{R}_3 \approx 35$ m;][]{2017ApJ...850L..36J}, this corresponds to a water-ice-covered fraction of the surface $\la 3\%$, which is relatively insensitive to the assumed input parameters. For comparison, such a low active fraction is also common for solar-system comets \citep[e.g.,][]{1995Icar..118..223A}. If the ice-covered fraction of the surface remained unchanged since around perihelion, at the time when the successful observations of `Oumuamua were conducted ($r_{\rm H} \ga 1.4$ au), the water production rate of `Oumuamua would be $Q \la \left(3.9 \pm 0.3\right) \times 10^{24}$ s$^{-1}$, which is smaller than all of the previously measured or estimated values by at least an order of magnitude \citep[c.f. ][and citations therein]{2019NatAs...3..594O}.

The motion of `Oumuamua has been identified to show a strong nongravitational effect \citep{2018Natur.559..223M}. The leading candidate causing the nongravitational acceleration is likely due to the mass-loss activity rather than the solar radiation pressure. To show why, we derive the ratio between the magnitude of the force due to mass loss ($F_{\rm ej}$) and the one due to the solar radiation pressure ($F_{\rm rad}$) as
\begin{equation}
\frac{F_{\rm ej}}{F_{\rm rad}} = \frac{\kappa c \left(1 + \mathcal{X} \right) \mathscr{U} \mathfrak{m}_{\rm H} Q v_{\rm th}}{\left(1+A_{\rm B}\right) S_{\odot} C_{\rm e}} \left( \frac{r_{\rm H}}{r_{\oplus}} \right)^{2}
\label{eq_f2f},
\end{equation}
\noindent where $c = 3 \times 10^8$ m s$^{-1}$ is the speed of light, $0 \le \kappa \le 1$ is the adimensional collimation efficiency of mass ejection, with the two ends corresponding to isotropic and perfectly collimated ejection, respectively, $\mathscr{U} = 18$ is the molecular weight of water, and $\mathfrak{m}_{\rm H} = 1.67 \times 10^{-27}$ kg is the mass of the hydrogen atom, and $v_{\rm th} = \sqrt{8 k_{\rm B} T / \left(\pi \mathscr{U} \mathfrak{m}_{\rm H}\right)}$ is the thermal speed. Here, $k_{\rm B} = 1.38 \times 10^{-23}$ J K$^{-1}$ is the Boltzmann constant. In Equation (\ref{eq_f2f}), we have assumed that the dust is well coupled with the gas drag. With $209 \le T \le 220$ K (solved from Equation (\ref{eq_E})) along with previously obtained and assumed values, we find that the solar radiation pressure force would be smaller than the counterpart due to the mass loss by at least three orders of magnitude. Therefore, assuming that the nongravitational acceleration arises from the mass loss of `Oumuamua seems more appropriate. 

We then write the momentum conservation equation for `Oumuamua:
\begin{equation}
\mathcal{M}_{\rm n} g\left(r_{\rm H}\right) \sqrt{\sum_{j=1}^{3} A_{j}^2} = \kappa\left(1 + \mathcal{X}\right) \mathscr{U} \mathfrak{m}_{\rm H} Q v_{\rm th}
\label{eq_mc},
\end{equation}
\noindent in which $\mathcal{M}_{\rm n}$ is the nucleus mass of `Oumuamua, $A_{j}$ ($j=1,2,3$) are respectively the radial, transverse and normal components of the nongravitational acceleration at $r_{\rm H} = 1$ au, and $g\left(r_{\rm H}\right)$ is a dimensionless mass-loss law with normalisation at $r_{\rm H} = 1$ au. The direction of the nongravitational acceleration of `Oumuamua is predominantly radial \citep{2018Natur.559..223M}. Thus, we can find $\rho_{\rm n}$ from Equation (\ref{eq_mc}) as
\begin{equation}
\rho_{\rm n} = \frac{3 \left(1 + \mathcal{X} \right) \kappa Q}{\pi A_{1} g\left(r_{\rm H} \right) \mathcal{R}_{1} \mathcal{R}_{2} \mathcal{R}_{3}} \sqrt{\frac{k_{\rm B} \mathscr{U} \mathfrak{m}_{\rm H} T}{2\pi}}
\label{eq_den_nuc}.
\end{equation}
\noindent With $A_1 = \left(2.8 \pm 0.4\right) \times 10^{-7}$ au d$^{-2}$, $g = 7.2$ (based on the JPL Horizons solution by D. Farnocchia), and the assumed $\mathcal{X} = 1$ at $r_{\rm H} = 0.375$ au, Equation (\ref{eq_den_nuc}) yields $\rho_{\rm n} \la 40$ kg m$^{-3}$ as the nucleus bulk density of `Oumuamua, which is not only smaller than the known values of rubble-pile asteroids \citep[e.g., $\rho_{\rm n} = \left(1.9 \pm 0.1\right) \times 10^{3}$ kg m$^{-3}$ for (25143) Itokawa;][]{2006Sci...312.1330F}, but also those of cometary nuclei in the solar system \citep[e.g., $\rho_{\rm n} = 533 \pm 6$ kg m$^{-3}$ for 67P/Churyumov–Gerasimenko;][]{2016Natur.530...63P} by more than an order of magnitude. In order to have the nucleus bulk density of `Oumuamua comparable to the ones of comets, we consider the following options:
\begin{enumerate}
\item `Oumuamua is an extremely dusty interstellar interloper with a larger value of dust-to-gas mass ratio $\mathcal{X} \ga 10$. While the majority of solar-system comets have $\mathcal{X} \la 2$ \citep{1992AJ....104..848S,1996A&AS..120..301S}, there do exist exceptions where comets are abnormally dust-rich \citep[$\mathcal{X} \ga 4$; ][]{1999AJ....117.1056J,2014ApJ...784L..23Y}. Comets 1P/Halley and C/1995 O1 (Hale-Bopp) even manifested $\mathcal{X} \ga 10$ in outbursts at $r_{\rm H} \ge 6$ au \citep{1992A&A...263..367S,1996A&A...314..957S}. Given these, we think that adopting large values of dust-to-gas mass ratio $\mathcal{X} \ga 10$ is a reasonable option.

\item The water production rate of `Oumuamua was orders of magnitude greater than the one predicted by the empirical function by \citet{2008LPICo1405.8046J}. We cannot reject this possibility, as none of the samples studied by \citet{2008LPICo1405.8046J} are as faint as `Oumuamua, and those faint members that appear to satisfy the correlation have no direct measurements of the water production rates, and thus we are uncertain whether the relationship still remains a good approximation in such a domain. 

\item The dust coma of `Oumuamua consisted of dominant dust grains that are unprecedentedly large compared to solar-system comets. In such a scenario, the forward-scattering enhancement would be less prominent at phase angle $\sim$170\degr, resulting in a higher upper limit to the reduced brightness of `Oumuamua. Dust grains of this size range, no longer coupled with the outflowing gas, have much lower speeds \citep[$\la$10 m s$^{-1}$ for centimetre-sized grains; ][]{2004come.book..265H}, such that $\mathcal{X}$ in Equation (\ref{eq_den_nuc}) should be dropped. Assuming that the empirical correlation by \citet{2008LPICo1405.8046J} is still applicable, we will then obtain a greater upper limit to the water production rate during the {\it STEREO} observations. Substituting gives a larger maximum allowed value of the nucleus bulk density by a factor of ten.

\end{enumerate}

To conclude, options 1 and 2 are both plausible to make the nucleus bulk density of `Oumuamua comparable to those of comets, although they likely indicate that `Oumuamua has some unusual but by no means outrageous properties in the context of solar-system comets. Option 3 requires that `Oumuamua is a peculiar object that perhaps is distinctly different from the known objects in our solar system.

The other possible way to solve the density issue is that `Oumuamua may have not yet begun the outgassing that lead to the exhibited nongravitational acceleration during the {\it STEREO} observations, rather the activity began at some point between these observations and the discovery. In the following, we will show that this scenario is plausible. The thermal conduction timescale of `Oumuamua is $\tau_{\rm c} \sim \left(\mathcal{R}_{1} \mathcal{R}_{2} \mathcal{R}_{3} \right)^{2/3} / \mathcal{K} \sim 10^2$-10$^3$ yr, where $\mathcal{K} \sim 10^{-7}$-10$^{-6}$ m$^{2}$ s$^{-1}$ is the typical range for planetary electric soils \citep[e.g., ][]{1996EM&P...72..185J}. We thus approximate the core temperature of `Oumuamua as the mean temperature from some initial epoch $t_{0}$ to the perihelion epoch $t_{\rm p}$, with $t_{\rm p} - t_{0} \ge 100$ yr, from
\begin{align}
\nonumber
T_{\rm c} & = \left[\frac{\left(1-A_{\rm B}\right) S_{\odot}r_{\oplus}^2}{4 \epsilon \sigma \left(t_{\rm p} - t_{0}\right)} \int\limits_{t_{\rm p}}^{t_{0}} \frac{\mathrm{d}t}{r_{\rm H}^{2} \left(t\right)}\right]^{1/4} \\
& \approx \left[\frac{\left(1-A_{\rm B}\right) S_{\odot}r_{\oplus}^2}{4 \epsilon \sigma \left(t_{\rm p} - t_{0}\right) \sqrt{\mu q \left(e+1\right)}} \left(\pi - \arccos \frac{1}{e}\right) \right]^{1/4}
\label{eq_T_core},
\end{align}
\noindent where $q = 0.256$ au and $e = 1.201$ are the perihelion distance and eccentricity of `Oumuamua, respectively, and $\mu = 1.327 \times 10^{20}$ m$^3$ s$^{-2}$ is the heliocentric gravitational constant. Equation (\ref{eq_T_core}) then yields $T_{\rm c} \la 80$ K as the temperature at the core of the nucleus. At perihelion, the surface temperature of an inactive nucleus of `Oumuamua would be $\ga$600 K, and likely even $\sim$800 K around the subsolar point, thereby forming an enormous temperature gradient of $\Delta T \ga 500$ K across the nucleus interior. This is completely consistent with the much more detailed thermal modelling works by \citet{2018NatAs...2..133F} and \citet{2018AJ....155..217S}. Assuming a nominal thermal expansion coefficient of $\alpha_{\rm v} \sim 10^{-6}$-10$^{-5}$ K$^{-1}$ and a Young's modulus of $Y \sim 10$-100 GPa, which are both typical for common rocks \citep[c.f.][and citations therein]{2010AJ....140.1519J,2012ApJ...757..127S}, we find that the thermal stress around the perihelion passage would be as large as $\sigma_{\rm th} = \alpha_{\rm v} Y \Delta T \ga 5$-500 MPa, which is most likely larger than the tensile strengths of cometary nuclei \citep[e.g., ][]{2004come.book..359P,2015A&A...583A..32G,2016SoSyR..50..225B}. Thus, it would not be surprising that `Oumuamua experienced thermal fracturing at some point between transiting the FOV of {\it STEREO} and discovery, whereby an inert surface layer was removed and preexisting subterranean volatiles were exposed to the sunlight.

\section{Summary}
\label{sec_sum}

We conclude our analysis of the {\it SOHO}/{\it STEREO} nondetections of interstellar interloper 1I/2017 U1 (`Oumuamua) as follows:

\begin{enumerate}

\item `Oumuamua was too faint to be visible in coronagraph images from {\it STEREO-A}'s COR2 (apparent {\it V}-band magnitude $m_{V} \ga 13.0 \pm 0.1$) and {\it SOHO}'s C3 ($m_V \ga 10.6$) around its perihelion passage.

\item Thanks to the forward-scattering viewing geometry, which potentially brightened a dust coma by two orders of magnitude, the {\it STEREO} nondetection offers us a tighter constraint on the physical properties of `Oumuamua. Around the observation, the effective cross-section of dust coma of `Oumuamua was $C_{\rm e} \la \left(2.1 \pm 0.2 \right) \times 10^{4}$ m$^2$ for an assumed geometric albedo of $p_V = 0.1$. If similar to typical solar-system comets, the total mass of the dust coma would then be $\mathcal{M}_{\rm d} \la 20 \pm 2$ kg, which renders unlikely the hypothesis by \citet{2019arXiv190108704S} that `Oumuamua had disrupted before the perihelion passage.

\item Accordingly, the water production rate is estimated to be $Q \la\left(6.1 \pm 0.5\right) \times 10^{25}$ s$^{-1}$. If scaled to the time when the successful observations were performed, its value would be smaller than all of the previous measurements by at least an order of magnitude.

\item For our assumptions of typical solar system comet behavior, this would imply that the active fraction of `Oumuamua was $\la 3\%$, which is not unusual compared to typical solar-system comets. Given an assumed dust-to-gas mass ratio value of $\mathcal{X} = 1$, the nucleus bulk density would then have to be as low as $\la$40 kg m$^{-3}$, smaller than the known values of cometary nuclei in the solar system by over an order of magnitude, to show the observed nongravitational effect in its motion.

\item The low nucleus bulk density can be reconciled with typical cometary values by altering our assumptions, including a higher dust-to-gas mass ratio, greater water production rate, or very large dust grains. We consider the first two plausible, but the third seems unlikely. Alternatively, `Oumuamua may have been only weakly active or inactive during the {\it STEREO} observations. In this scenario, the outgassing needed to account for the observed nongravitational acceleration likely began postperihelion, and was potentially initiated by the removal of an inert surface layer due to the overwhelming thermal stress built up in its nucleus interior during the perihelion passage.

\end{enumerate}

\acknowledgements
{
We thank the anonymous referee for a speedy review and insightful comments, Brian Skiff  for discussions on photometric calibrations, Jian-Yang Li and Ludmilla Kolokolova for discussions on light scattering properties of dust. This paper benefited from useful discussions with the International Space Science Institute (ISSI) `Oumuamua team in Bern, Switzerland. The {\it SOHO}/LASCO data used here are produced by a consortium of the NRL (USA), Max-Planck-Institut fuer Aeronomie (Germany), Laboratoire d'Astronomie (France), and the University of Birmingham (UK). The {\it STEREO}/SECCHI project is an international consortium of the NRL, LMSAL and NASA GSFC (USA), RAL and University of Birmingham (UK), MPS (Germany), CSL (Belgium), IOTA and IAS (France). This work was funded by NASA Near Earth Object Observations grant No. NNX17AK15G and NASA Solar System Workings grant No. 80NSSC19K0024 to M.M.K.
}

\vspace{5mm}
\facilities{{\it SOHO}, {\it STEREO}}

\software{IDL, SolarSoftWare \citep{1998SoPh..182..497F}, AstroImageJ \citep{2013ascl.soft09001C}}


\end{document}